\documentclass[aps,prl,reprint,twocolumn,superscriptaddress,showpacs,nofootinbib]{revtex4-1}

\usepackage{float}
\usepackage{amsfonts}
\usepackage{amsmath}
\usepackage{amssymb}
\usepackage{latexsym}
\usepackage{array}
\usepackage{multirow}
\usepackage{ifpdf}
\usepackage{esint}
\usepackage{dsfont}
\usepackage{slashed}
\usepackage{color}
\usepackage{soul}

\ifpdf
\usepackage[pdftex]{graphicx}
\else
\usepackage{graphicx}
\fi

\ifpdf \DeclareGraphicsExtensions{.pdf, .jpg, .tif, .png} \else
\DeclareGraphicsExtensions{.eps, .jpg} \fi

\newcommand{\D}{\mathrm{d}}

\newcommand{\dB}{\mathrm{dB}}
\newcommand{\npsi}{\varPsi}
\newcommand{\tz}{\tilde{z}}
\newcommand{\tF}{\tilde{F}}

\newcommand{\tr}{\tilde{r}}
\newcommand{\ttt}{\tilde{t}}
\newcommand{\tpsi}{\tilde{\varPsi}}

\newcommand{\hz}{\hat{z}}
\newcommand{\hF}{\hat{F}}

\newcommand{\hpsi}{\hat{\varPsi}}
\newcommand{\bz}{\mathbf{z}}
\newcommand{\bF}{\mathbf{F}}
\newcommand{\bE}{\mathbf{E}}
\newcommand{\bV}{\mathbf{V}}

\newcommand{\cW}{\mathcal{W}}

\begin{document}

\title{Quantum reflection and Liouville transformations from wells to walls}

\author{G. Dufour} \email[]{gabriel.dufour@upmc.fr}
\affiliation{Laboratoire Kastler Brossel, UPMC-Sorbonne
Universit\'es, CNRS, ENS-PSL Research University, Coll\`ege de
France, Campus Jussieu, F-75252 Paris, France.}
\author{R. Gu\'erout}
\affiliation{Laboratoire Kastler Brossel, UPMC-Sorbonne
Universit\'es, CNRS, ENS-PSL Research University, Coll\`ege de
France, Campus Jussieu, F-75252 Paris, France.}
\author{A. Lambrecht}
\affiliation{Laboratoire Kastler Brossel, UPMC-Sorbonne
Universit\'es, CNRS, ENS-PSL Research University, Coll\`ege de
France, Campus Jussieu, F-75252 Paris, France.}
\author{S. Reynaud}
\affiliation{Laboratoire Kastler Brossel, UPMC-Sorbonne
Universit\'es, CNRS, ENS-PSL Research University, Coll\`ege de
France, Campus Jussieu, F-75252 Paris, France.}
\date{\today}

\begin{abstract}
Liouville transformations map in a rigorous manner one Schr\"odinger
equation into another, with a changed scattering potential. They are
used here to transform quantum reflection of an atom on an
attractive \emph{well} into reflection of the atom on a repulsive
\emph{wall}. While the scattering properties are preserved, the
corresponding semiclassical descriptions are completely different. A
quantitative evaluation of quantum reflection probabilities is
deduced from this method.
\end{abstract}

\maketitle

Quantum reflection of atoms from the van der Waals attraction to a
surface has been studied theoretically since the early days of
quantum mechanics \cite{Lennard-Jones1936,Lennard-Jones1936a}.
Though the classical motion would be increasingly accelerated
towards the surface, the quantum matter waves are reflected back
with a probability that approaches unity at low energies, because
the potential varies more and more rapidly close to the surface.
Experiments have seen quantum reflection for He and H atoms on
liquid helium films \cite{Nayak1983,Berkhout1989,Yu1993} and for
ultracold atoms or molecules on solid surfaces
\cite{Shimizu2001,Druzhinina2003,Pasquini2004,Oberst2005,Pasquini2006,%
Zhao2010,Zhao2011}. Meanwhile various fundamental aspects and
applications have been analyzed in a number of theoretical papers
\cite{Berry1972,Boheim1982,Clougherty1992,Carraro1992,Henkel1996,%
Friedrich2002,Friedrich2004,Friedrich2004a,Voronin2005,Judd2011,Voronin2012}.

Paradoxical phenomena appear in the study of quantum reflection from
the Casimir-Polder (CP) interaction with a surface. The potential is
attractive, with characteristic inverse power laws at both ends of
the physical domain $z\in\ ]0,\infty[$ delimited by the material
surface located at $z=0$ : $V(z)\simeq-C_3/z^3$ at the
\emph{cliff-side}, close to the surface and $V(z)\simeq-C_4/z^4$ at
the \emph{far-end}, away from it. Strikingly, the probability of
reflection increases when the energy $E$ of the incident atom is
decreased, and increases as well when the absolute magnitude of the
potential is decreased. For example, the probability of quantum
reflection is larger for atoms falling onto silica bulk than onto
metallic or silicon bulks \cite{Dufour2013qrefl} and is even larger
for nanoporous silica \cite{Dufour2013porous}.

In the present letter, we use Liouville transformations to study
quantum reflection (QR). In quantum mechanics, a Liouville
transformation maps in a rigorous manner one Schr\"odinger equation
into another, with a changed scattering potential. In a
semiclassical picture however, the problem can be transformed from
QR of an atom on an attractive \emph{well} into a problem of
reflection on a repulsive \emph{wall}. Remarkably, scattering
properties are invariant under the Liouville transformation and the
paradoxical features of the initial QR problem become intuitive
predictions of the better defined problem of reflection on the
repulsive wall. We will also obtain a quantitative evaluation of QR
probabilities in this way.

We consider a cold atom of mass $m$ incident with an energy $E>0$ on
the CP potential $V(z)$ in the half-line $z\in\ ]0,\infty[$. For
plane material surfaces, the motion orthogonal to the plane (along
the $z-$direction) is decoupled from the transverse motions and
described by a 1D Schr\"odinger equation:
\begin{eqnarray}
\label{schrod} \npsi^{\prime\prime}(z) + F(z) \, \npsi(z) = 0 \,,\,
F(z)\equiv\frac{2m\left(E-V(z)\right)}{\hbar^2}~.&&
\end{eqnarray}
Throughout the letter, primes denote differentiation with respect to
the argument of the function.

In the semiclassical WKB approximation, the function $F(z)$ is seen
as the square of the de Broglie wave-vector $k_\dB$ associated with
the classical momentum $p\equiv\hbar k_\dB$. As the CP potential is
attractive and the incident energy positive, $F$ is positive, so
that a classical particle undergoes an increasing acceleration
towards the surface. For a quantum particle in contrast, QR occurs
when the variation of $k_\dB$ becomes significant on a length scale
of the order of the de Broglie wavelength:
\begin{eqnarray} \label{lambdadB} \lambdabar_\dB \equiv
\frac{\lambda_\dB}{2\pi} \equiv \frac1{k_\dB} = \frac1{\sqrt{F}}
~.&&
\end{eqnarray}

The Schr\"odinger equation \eqref{schrod} can be solved in full
generality by writing its solution as a linear combination of
counter-propagating WKB waves with $z-$dependent coefficients and
matching it to the appropriate boundary conditions at both ends of
the physical domain \cite{Berry1972}. Matter-waves can be reflected
back from the cliff-side so that the complete problem depends on the
details of the physics of the surface. In this letter, we focus our
attention on the one-way problem where the CP potential is crossed
only once and, therefore, do not discuss this surface physics
problem any longer. The numerical solution of \eqref{schrod} leads
to reflection and transmission amplitudes depending on the incident
energy $E$ or, equivalently, of the parameter $\kappa \equiv
\sqrt{2mE}/\hbar$ which is also the asymptotic value of de Broglie
wavevector in the far-end.

In spite of its effectiveness, the numerical solution of the QR
problem leaves open questions. First, the scattering problem, where
matter waves are reflected or transmitted on the CP potential, is
not well defined with the potential diverging at the cliff-side.
Second, an intuitive understanding of the dependence of QR
probability on the parameters is missing. The Liouville
transformations considered in the following will give clear answers
to these questions.

The Schr\"odinger equation \eqref{schrod} is an example of a
Sturm-Liouville equation in canonical form \cite{Liouville1836},
which can be submitted to transformations introduced by Liouville
\cite{Liouville1837} and often named after him (see the historical
notes at the end of ch.6 in \cite{Olver1997}). We stress at this
point that we use these transformations to relate exactly equivalent
scattering problems, with no approximation (see a similar approach
to the study of differential equations in \cite{Milson1998}).

Liouville transformations are gauge transformations consisting in a
change of coordinate $z \to \tz$, with $\tz(z)$ a smooth
monotonously increasing function, and an associated rescaling of the
wave-function:
\begin{eqnarray} \label{liouville}
\tpsi(\tz)=\sqrt{\tz'(z)}\,\npsi(z) ~.
\end{eqnarray}
Equation \eqref{schrod} for $\npsi$ is transformed under
\eqref{liouville} into an equivalent equation for $\tpsi$ with
\cite{Olver1997}:
\begin{eqnarray}
\label{transformed}
\tF(\tz)=\frac{F(z)-\tfrac{1}{2}\{\tz,z\}}{\tz'(z)^2} = z'(\tz)^2
F(z) + \tfrac{1}{2} \{z,\tz\}~.
\end{eqnarray}
The curly braces denote the Schwarzian derivative of the
coordinate transformation:
\begin{eqnarray}
\label{schwarz} \{\tz,z\} = \frac{\tz'''(z)}{\tz'(z)} -
\frac{3}{2}\frac{\tz''(z)^2}{\tz'(z)^2} ~.&&
\end{eqnarray}

These transformations form a group, with the composition of $z\to
\tz$ and $\tz\to\hz$ being a transformation $z\to \hz$. The
compatibility of relations obeyed by $(\npsi,F)$, $(\tpsi,\tF)$ and
$(\hpsi,\hF)$ is ensured by Cayley's identity:
\begin{eqnarray} \label{cayley}
\left\{\hz,z\right\} =
\left(\tz'(z)\right)^{2}\,\left\{\hz,\tz\right\} +
\left\{\tz,z\right\}.
\end{eqnarray}
The inverse transformation, used for the second equality in
\eqref{transformed}, is obtained by applying \eqref{cayley} to the
case $\hz=z$.

The group of transformations preserves the Wronskian of two
solutions $\npsi_1,\npsi_2$ of the Schr\"odinger equation, which is
a constant independent of $z$ and skew symmetric in the exchange of
the two solutions:
\begin{eqnarray}
\label{wronskian} &&\cW\left(\npsi_1,\npsi_2\right) =
\npsi_1(z)\npsi_2'(z) - \npsi_1'(z)\npsi_2(z) ~.
\end{eqnarray}
In particular, when $\npsi$ solves \eqref{schrod}, its complex
conjugate $\npsi^*$ solves it as well. As the probability density
current is proportional to the Wronskian  $\cW(\npsi^*,\npsi)$, it
is invariant under the transformation. The reflection and
transmission amplitudes $r$ and $t$ are also preserved, as they can
be written in terms of Wronskians of solutions which match incoming
and outgoing WKB waves \cite{Whitton1973}. They can be calculated
equivalently after any Liouville transformation, with $\tr=r$ and
$\ttt=t$. These transformations, which do not necessarily simplify
the resolution of \eqref{schrod}, have to be considered as gauge
transformations relating equivalent scattering problems to one
another.

These quantum-mechanically equivalent scattering problems may
correspond to extremely different classical descriptions. We now
write a specific Liouville gauge which maps the initial problem of
QR on an attractive well into an intuitively different problem of
reflection on a repulsive wall. This choice brings clear answers to
the questions discussed above, and it will allow us to uncover
scaling relations between the QR probabilities and the parameters of
the problem.

This specific Liouville gauge is written in terms of the WKB phase
$\phi\equiv\int^z k_\dB(y) \D y$ associated with the classical
action integral $S\equiv\hbar\phi$. We fix the freedom associated
with the arbitrariness of the phase reference by enforcing $\phi(z)
\to \kappa z$ at $z\to\infty$. We then choose the coordinate $\bz$
for which we get quantities identified by boldfacing:
\begin{eqnarray} \label{specific} && \bz\equiv
\frac\phi{\sqrt{\kappa\ell}} ~,
\quad \bF(\bz) \equiv \bE - \bV(\bz) ~, \\
&&\bE = \kappa\ell ~, \quad \bV(\bz) = -\kappa\ell
\sqrt{\lambdabar_\dB^3(z)}
\left(\sqrt{\lambdabar_\dB(z)}\right)^{\prime\prime}  \nonumber ~.
\end{eqnarray}
We have defined the length scale $\ell \equiv \sqrt{2mC_4}/\hbar$
associated with the far-end tail of the CP potential. Its
introduction in \eqref{specific} has been done for reasons which
will become clear soon, and it leads to a dimensionless energy $\bE$
and a dimensionless potential $\bV$.

For the CP potential, the quantity $\bV$ vanishes at both ends of
the physical domain $z\in\,]0,\infty[$, that is also at both ends of
the transformed domain $\bz\in\,]-\infty,\infty[$, so that the
problem now corresponds to a well-defined scattering problem with no
interaction in the asymptotic input and output states. In striking
contrast with the original QR problem, the transformed problem can
have classical turning points where $\bF=0$ or $\bE=\bV$, though it
corresponds to the same scattering amplitudes.

\begin{figure}[th]
    \includegraphics[width=0.5\textwidth]{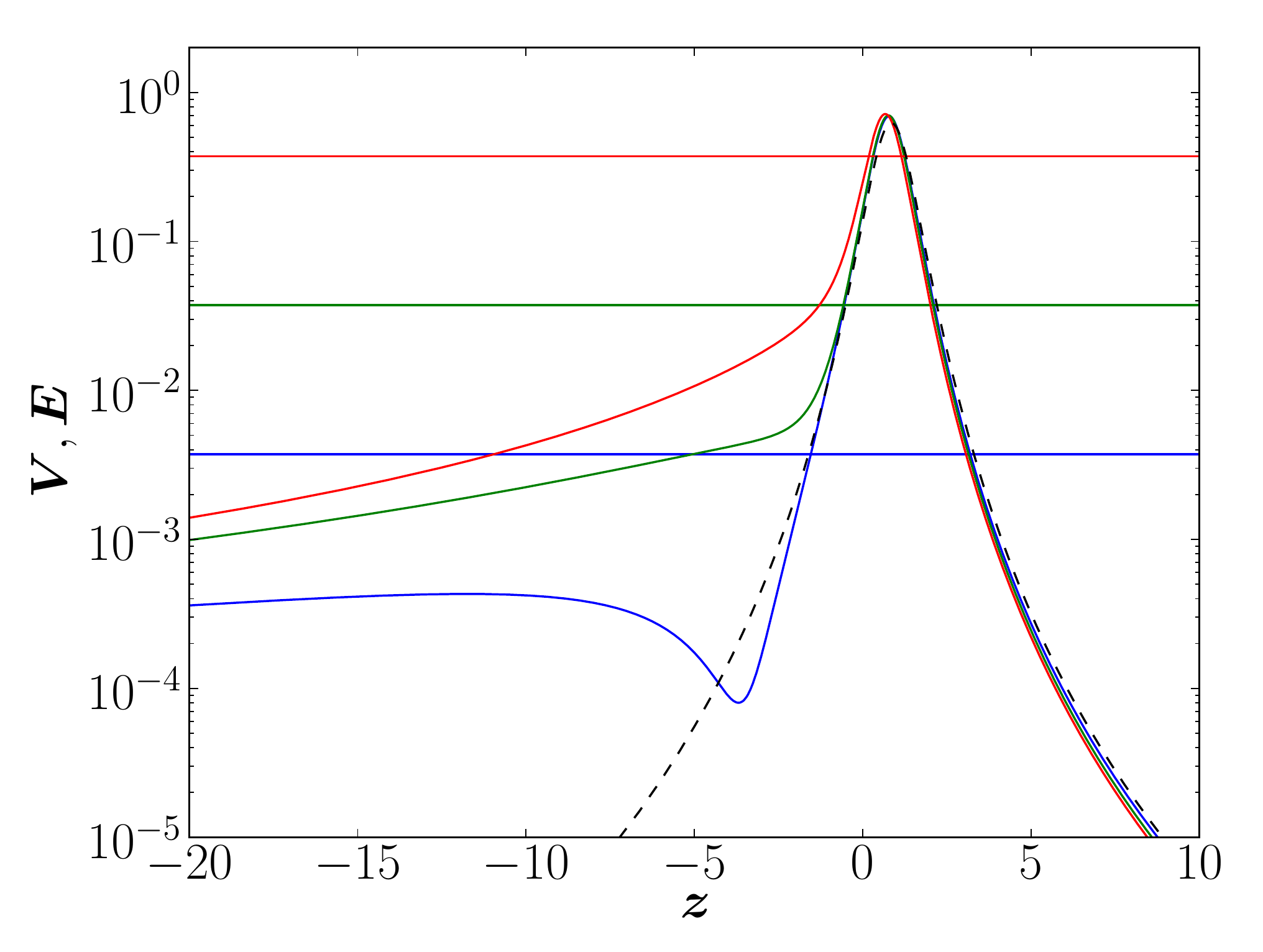}
\caption{[Colors online] The plots represent the constants $\bE$
(horizontal lines) and the functions $\bV(\bz)$ (curves) calculated
for different scattering problems, corresponding to the same CP
potential $V(z)$ between an hydrogen atom and a silica bulk and
energies $E$ equal to 0.001, 0.1 and 10 neV (respectively blue,
green and red from the lowest to the highest value of $\bE$, or from
the lowest to the highest value of $\bV$ in the left-hand part of
the plot). The dashed (black) curve is the universal function
$\bV(\bz)$ calculated for a pure $C_4$ model.
\label{fig:silica-energy} }
\end{figure}

This important point is illustrated by the drawings on
Fig.\ref{fig:silica-energy}, which shows the constants $\bE$ and the
functions $\bV(\bz)$ for different scattering problems. In all
cases, the original potential $V$ is calculated for the CP
interaction between an hydrogen atom and a silica bulk
\cite{Dufour2013qrefl}, whereas the incident energies $E$ are
respectively equal to 0.001, 0.1 and 10 neV. With $\bE$ always
positive and $\bV(\bz)$ often positive, a logarithmic scale is used
along the vertical axis, which makes some details more apparent.

The most striking feature of these plots is the appearance of
classical turning points for the not too high energies considered
here, so that QR on an attractive well is now intuitively understood
as reflection on a wall. Other clearly visible properties are that
$\bE$ scales like $\sqrt{E}$ whereas $\bV(\bz)$ has nearly identical
peak shapes for different energies. The fact that the QR probability
goes to unity when $E\to0$ is now an immediate consequence of the
increasing reflection expected for a particle with a decreasing
energy $\bE$ coming onto a wall with a peak $\bV$.

In fact, the potentials $\bV$ calculated for different energies tend
to build up a universal function at large enough values of $\bz$,
and this universal function has a symmetrical shape. These two facts
can be explained by looking at the particular model $V(z)=-C_4/z^4$,
which is representative of the CP interaction in the far-end. For
this simple model, $\bV(\bz)$ is given by parametric relations (with
$e^u \equiv z/z_0$ and $z_0=\sqrt{\ell/\kappa}$):
\begin{eqnarray}
\label{universal} &&\bV = \frac5{8\cosh^3(2u)} ~,\\
&&\bz = \bz_0+\int^u_0 \sqrt{2\cosh(2v)} \D v ~,\quad \bz_0
=\frac{1}{\sqrt{\pi}}\Gamma\!\left(\tfrac34\right)^2  ~.\nonumber
\end{eqnarray}
This function, drawn as the dashed curve on
Fig.\ref{fig:silica-energy}, reaches its peak value $\tfrac58$ at
$z=z_0$, which lies further and further away from the surface when
the energy decreases. This also explains why the functions plotted
on Fig.\ref{fig:silica-energy} for the full CP potential tend to
align on this universal form when the energy decreases. The
deviations appearing on the figure correspond to values of $z$ near
the cliff-side, for which the $C_4$ model is indeed a poor
representation as the potential behaves as $-C_3/z^3$. In the
parametric definition \eqref{universal}, $\bV$ is even and
$\bz-\bz_0$ odd in the parity $u\to-u$. It follows that the
universal function $\bV(\bz)$ is symmetrical with respect to
$\bz_0$.

\begin{figure}[bh]
  \begin{center}
   \includegraphics[width=0.5\textwidth]{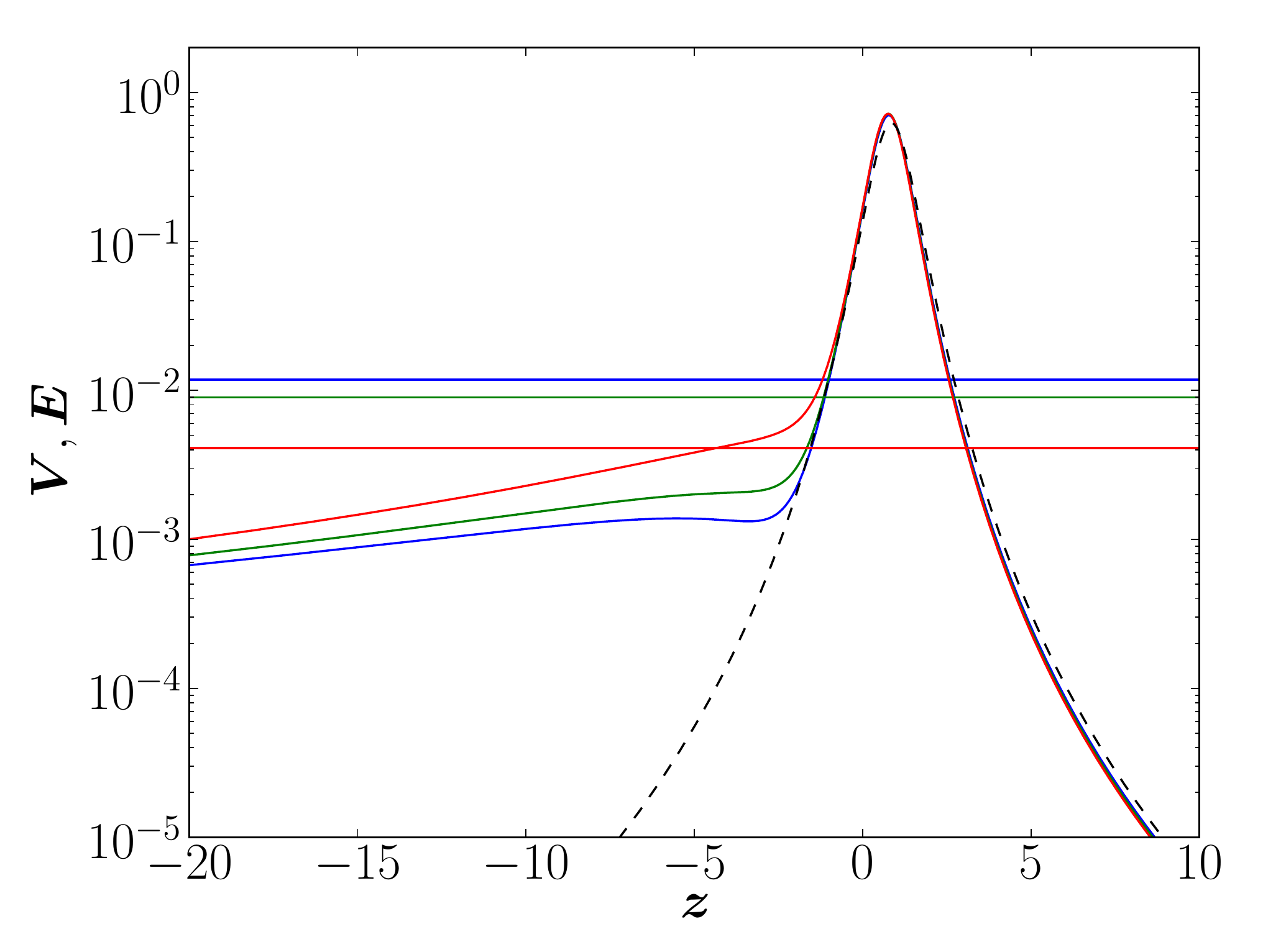}
  \end{center}
\caption{[Colors online] The plots represent the constants $\bE$
(horizontal lines) and the functions $\bV(\bz)$ (curves) calculated
for different scattering problems, corresponding to the energy
$E=0.01$ neV and the CP potentials $V(z)$ between an hydrogen atom
and nanoporous silica with porosities $\eta$ equal to 0\%, 50\% and
90\% (respectively blue, green and red from the highest to the
lowest value of $\bE$, or from the lowest to the highest value of
$\bV$ in the left-hand part of the plot). The dashed (black) curve
is the same as on Fig.\ref{fig:silica-energy}.
\label{fig:silica-porosity} }
\end{figure}

We come now to the discussion of the dependence of QR on the
absolute magnitude of the CP potential. To do so we consider
hydrogen falling onto nanoporous silica, which has a weaker CP
interaction when its porosity increases \cite{Dufour2013porous}.
Fig.\ref{fig:silica-porosity} shows the constants $\bE$ and the
functions $\bV(\bz)$ for an energy $E=0.01$~neV, and the potentials
calculated for an hydrogen atom falling onto nanoporous silica with
porosities $\eta$ equal to 0\%, 50\% and 90\%. These potentials
correspond to different far-end tails, with values of $C_4$, and
therefore $\ell$, smaller and smaller when the porosity is
increased. As on Fig.\ref{fig:silica-energy}, the transformed
potentials $\bV$ have nearly identical peak shapes, which tend to
align on the universal curve calculated for a pure $C_4$ potential
and shown as the dashed curve. In contrast, the transformed energies
$\bE=\kappa\ell$ are decreasing when $\ell$ is decreased, which
immediately explains why the QR probability increases
\cite{Dufour2013porous}.

We finally discuss the values obtained for QR probabilities, by
comparing the exact results for the full CP potential with those
obtained for the $C_4$ model. To this aim, we first recall the low
energy behavior of the QR probability:
\begin{eqnarray}
\label{defb} R(\kappa) \equiv \vert r(\kappa) \vert^2 \simeq 1-4
\kappa b ~, \quad \kappa\to0 ~,&&
\end{eqnarray}
where $b$ is the opposite of the imaginary part of the scattering
length \cite{Dufour2013qrefl}. For a pure $C_4$ model, $b$ is known
to be equal to $\ell$ \cite{Voronin2005}, but this is not the case
for the full CP potential. Table \ref{bandell} gives $\ell$ and $b$
for nanoporous silica with different porosities $\eta$ ($\eta=0$\%
for silica bulk).

\begin{table}[h]
\begin{center}
\begin{tabular}{|c|c|c|c|c|c|}
\hline $\eta$ [\%]& 0 & 30 & 50 & 70 & 90  \\
\hline $\ell$ [$a_0$] &321.3 &  282.1 & 244.7 & 192.8 &  111.8  \\
\hline $b$ [$a_0$] & 272.7 & 227.8 &  187.5 &  134.0 & 57.0  \\
\hline
\end{tabular}
\caption{Values of $\ell$ and $b$ calculated for different
porosities, measured in atomic units $a_0\simeq53$~pm.
\label{bandell} }
\end{center}
\end{table}

\begin{figure}[h]
  \begin{center}
   \includegraphics[width=0.5\textwidth]{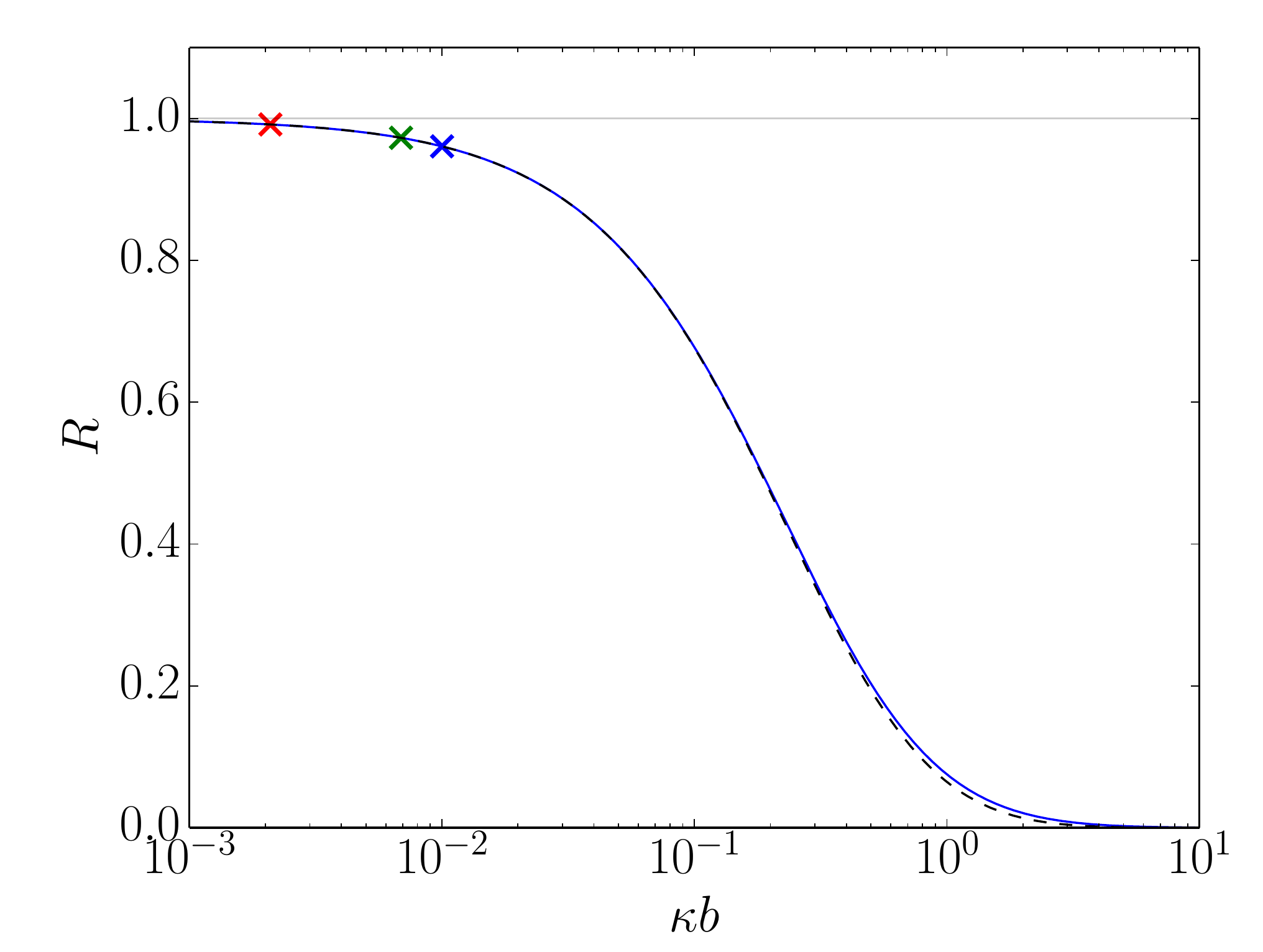}
  \end{center}
\caption{[Colors online] Quantum reflection probability $R$ shown as
a function of the dimensionless parameter $\kappa b$. The full blue
curve represents the values calculated for silica bulks in
\cite{Dufour2013qrefl}, while the crosses correspond to the
scattering problems of Fig.\ref{fig:silica-porosity} with the same
color codes. The dashed (black) curve is the universal function $R$
for a pure $C_4$ model. \label{fig:Rvskb}}
\end{figure}

We have reported on Fig.\ref{fig:Rvskb} the calculated QR
probabilities $R$ as a function of the dimensionless parameter
$\kappa b$ for the scattering problems discussed above. The full
blue curve represents the values calculated for silica bulks in
\cite{Dufour2013qrefl}, while the circles correspond to the
scattering problems of Fig.\ref{fig:silica-porosity} with the same
color code. The dashed black curve corresponds to the universal
function $R\left(\kappa b\right)$ obtained for the pure $C_4$ model,
with $b\equiv\ell$ in this case. A table of values of this function
is available as supplemental material. The exact results on
Fig.\ref{fig:Rvskb} are hardly distinguishable from this universal
function, except at large values of $\kappa b$ where QR
probabilities are small anyway.

In this letter, the problem of QR of an atom on a potential well has
been mapped into an equivalent problem of reflection on a wall
through a Liouville transformation of the Schrödinger equation. This
exact transformation relates quantum scattering processes which
correspond to different semiclassical pictures. It produces a new
and clear interpretation of the main features of quantum reflection
which were counterintuitive in the initial problem. It also allows
quantitative evaluation of QR probabilities which can be obtained
from the universal function corresponding to the pure $C_4$ model.

\textit{Acknowledgements - } Thanks are due for insightful
discussions to M.-T. Jaekel, V.V. Nesvizhevsky, A. Yu. Voronin, and
the GBAR and GRANIT collaborations.

\bibliography{bibliography}

\begin{thebibliography}{32}%
\makeatletter
\providecommand \@ifxundefined [1]{%
 \@ifx{#1\undefined}
}%
\providecommand \@ifnum [1]{%
 \ifnum #1\expandafter \@firstoftwo
 \else \expandafter \@secondoftwo
 \fi
}%
\providecommand \@ifx [1]{%
 \ifx #1\expandafter \@firstoftwo
 \else \expandafter \@secondoftwo
 \fi
}%
\providecommand \natexlab [1]{#1}%
\providecommand \enquote  [1]{``#1''}%
\providecommand \bibnamefont  [1]{#1}%
\providecommand \bibfnamefont [1]{#1}%
\providecommand \citenamefont [1]{#1}%
\providecommand \href@noop [0]{\@secondoftwo}%
\providecommand \href [0]{\begingroup \@sanitize@url \@href}%
\providecommand \@href[1]{\@@startlink{#1}\@@href}%
\providecommand \@@href[1]{\endgroup#1\@@endlink}%
\providecommand \@sanitize@url [0]{\catcode `\\12\catcode `\$12\catcode
  `\&12\catcode `\#12\catcode `\^12\catcode `\_12\catcode `\%12\relax}%
\providecommand \@@startlink[1]{}%
\providecommand \@@endlink[0]{}%
\providecommand \url  [0]{\begingroup\@sanitize@url \@url }%
\providecommand \@url [1]{\endgroup\@href {#1}{\urlprefix }}%
\providecommand \urlprefix  [0]{URL }%
\providecommand \Eprint [0]{\href }%
\@ifxundefined \urlstyle {%
  \providecommand \doi  [0]{\begingroup \@sanitize@url \@doi}%
  \providecommand \@doi [1]{\endgroup \@@startlink {\doibase
  #1}doi:\discretionary {}{}{}#1\@@endlink }%
}{%
  \providecommand \doi  [0]{doi:\discretionary{}{}{}\begingroup
  \urlstyle{rm}\Url }%
}%
\providecommand \doibase [0]{http://dx.doi.org/}%
\providecommand \Doi [0]{\begingroup \@sanitize@url \@Doi }%
\providecommand \@Doi  [1]{\endgroup\@@startlink{\doibase#1}\@@Doi}%
\providecommand \@@Doi [1]{#1\@@endlink}%
\providecommand \selectlanguage [0]{\@gobble}%
\providecommand \bibinfo  [0]{\@secondoftwo}%
\providecommand \bibfield  [0]{\@secondoftwo}%
\providecommand \translation [1]{[#1]}%
\providecommand \BibitemOpen [0]{}%
\providecommand \bibitemStop [0]{}%
\providecommand \bibitemNoStop [0]{.\EOS\space}%
\providecommand \EOS [0]{\spacefactor3000\relax}%
\providecommand \BibitemShut  [1]{\csname bibitem#1\endcsname}%
\bibitem [{\citenamefont {Lennard-Jones}\ and\ \citenamefont
  {Devonshire}(1936){\natexlab{a}}}]{Lennard-Jones1936}%
  \BibitemOpen
  \bibfield  {author} {\bibinfo {author} {\bibfnamefont {J.~E.}\ \bibnamefont
  {Lennard-Jones}}\ and\ \bibinfo {author} {\bibfnamefont {A.~F.}\ \bibnamefont
  {Devonshire}},\ }\Doi {\href{10.1098/rspa.1936.0131}{10.1098/rspa.1936.0131}}
  {\bibfield  {journal} {\bibinfo  {journal} {{Proc. Roy. Soc. London}},\
  }\textbf {\bibinfo {volume} {{A156}}},\ \bibinfo {pages} {6} (\bibinfo {year}
  {1936}{\natexlab{a}})}\BibitemShut {NoStop}%
\bibitem [{\citenamefont {Lennard-Jones}\ and\ \citenamefont
  {Devonshire}(1936){\natexlab{b}}}]{Lennard-Jones1936a}%
  \BibitemOpen
  \bibfield  {author} {\bibinfo {author} {\bibfnamefont {J.~E.}\ \bibnamefont
  {Lennard-Jones}}\ and\ \bibinfo {author} {\bibfnamefont {A.~F.}\ \bibnamefont
  {Devonshire}},\ }\Doi {\href{10.1098/rspa.1936.0132}{10.1098/rspa.1936.0132}}
  {\bibfield  {journal} {\bibinfo  {journal} {{Proc. Roy. Soc. London}},\
  }\textbf {\bibinfo {volume} {{A156}}},\ \bibinfo {pages} {29} (\bibinfo {year}
  {1936}{\natexlab{b}})}\BibitemShut {NoStop}%
\bibitem [{\citenamefont {Nayak}\ \emph {et~al.}(1983)\citenamefont {Nayak},
  \citenamefont {Edwards},\ and\ \citenamefont {Masuhara}}]{Nayak1983}%
  \BibitemOpen
  \bibfield  {author} {\bibinfo {author} {\bibfnamefont {V.~U.}\ \bibnamefont
  {Nayak}}, \bibinfo {author} {\bibfnamefont {D.~O.}\ \bibnamefont {Edwards}},
  \ and\ \bibinfo {author} {\bibfnamefont {N.}~\bibnamefont {Masuhara}},\ }\Doi
  {\href{10.1103/PhysRevLett.50.990}{10.1103/PhysRevLett.50.990}} {\bibfield
  {journal} {\bibinfo  {journal} {{Phys. Rev. Lett.}},\ }\textbf {\bibinfo
  {volume} {{50}}},\ \bibinfo {pages} {990} (\bibinfo {year}
  {1983})}\BibitemShut {NoStop}%
\bibitem [{\citenamefont {Berkhout}\ \emph {et~al.}(1989)\citenamefont
  {Berkhout}, \citenamefont {Luiten}, \citenamefont {Setija}, \citenamefont
  {Hijmans}, \citenamefont {Mizusaki},\ and\ \citenamefont
  {Walraven}}]{Berkhout1989}%
  \BibitemOpen
  \bibfield  {author} {\bibinfo {author} {\bibfnamefont {J.~J.}\ \bibnamefont
  {Berkhout}}, \bibinfo {author} {\bibfnamefont {O.~J.}\ \bibnamefont
  {Luiten}}, \bibinfo {author} {\bibfnamefont {I.~D.}\ \bibnamefont {Setija}},
  \bibinfo {author} {\bibfnamefont {T.~W.}\ \bibnamefont {Hijmans}}, \bibinfo
  {author} {\bibfnamefont {T.}~\bibnamefont {Mizusaki}}, \ and\ \bibinfo
  {author} {\bibfnamefont {J.~T.~M.}\ \bibnamefont {Walraven}},\ }\Doi
  {\href{10.1103/PhysRevLett.63.1689}{10.1103/PhysRevLett.63.1689}} {\bibfield
  {journal} {\bibinfo  {journal} {{Phys. Rev. Lett.}},\ }\textbf {\bibinfo
  {volume} {{63}}},\ \bibinfo {pages} {1689} (\bibinfo {year}
  {1989})}\BibitemShut {NoStop}%
\bibitem [{\citenamefont {Yu}\ \emph {et~al.}(1993)\citenamefont {Yu},
  \citenamefont {Doyle}, \citenamefont {Sandberg}, \citenamefont {Cesar},
  \citenamefont {Kleppner},\ and\ \citenamefont {Greytak}}]{Yu1993}%
  \BibitemOpen
  \bibfield  {author} {\bibinfo {author} {\bibfnamefont {I.~A.}\ \bibnamefont
  {Yu}}, \bibinfo {author} {\bibfnamefont {J.~M.}\ \bibnamefont {Doyle}},
  \bibinfo {author} {\bibfnamefont {J.~C.}\ \bibnamefont {Sandberg}}, \bibinfo
  {author} {\bibfnamefont {C.~L.}\ \bibnamefont {Cesar}}, \bibinfo {author}
  {\bibfnamefont {D.}~\bibnamefont {Kleppner}}, \ and\ \bibinfo {author}
  {\bibfnamefont {T.~J.}\ \bibnamefont {Greytak}},\ }\Doi
  {\href{10.1103/PhysRevLett.71.1589}{10.1103/PhysRevLett.71.1589}} {\bibfield
  {journal} {\bibinfo  {journal} {{Phys. Rev. Lett.}},\ }\textbf {\bibinfo
  {volume} {{71}}},\ \bibinfo {pages} {1589} (\bibinfo {year}
  {1993})}\BibitemShut {NoStop}%
\bibitem [{\citenamefont {Shimizu}(2001)}]{Shimizu2001}%
  \BibitemOpen
  \bibfield  {author} {\bibinfo {author} {\bibfnamefont {F.}~\bibnamefont
  {Shimizu}},\ }\Doi
  {\href{10.1103/PhysRevLett.86.987}{10.1103/PhysRevLett.86.987}} {\bibfield
  {journal} {\bibinfo  {journal} {{Phys. Rev. Lett.}},\ }\textbf {\bibinfo
  {volume} {{86}}},\ \bibinfo {pages} {987} (\bibinfo {year}
  {2001})}\BibitemShut {NoStop}%
\bibitem [{\citenamefont {Druzhinina}\ and\ \citenamefont
  {DeKieviet}(2003)}]{Druzhinina2003}%
  \BibitemOpen
  \bibfield  {author} {\bibinfo {author} {\bibfnamefont {V.}~\bibnamefont
  {Druzhinina}}\ and\ \bibinfo {author} {\bibfnamefont {M.}~\bibnamefont
  {DeKieviet}},\ }\Doi
  {\href{10.1103/PhysRevLett.91.193202}{10.1103/PhysRevLett.91.193202}}
  {\bibfield  {journal} {\bibinfo  {journal} {{Phys. Rev. Lett.}},\ }\textbf
  {\bibinfo {volume} {{91}}},\ \bibinfo {pages} {193202} (\bibinfo {year}
  {2003})}\BibitemShut {NoStop}%
\bibitem [{\citenamefont {Pasquini}\ \emph {et~al.}(2004)\citenamefont
  {Pasquini}, \citenamefont {Shin}, \citenamefont {Sanner}, \citenamefont
  {Saba}, \citenamefont {Schirotzek}, \citenamefont {Pritchard},\ and\
  \citenamefont {Ketterle}}]{Pasquini2004}%
  \BibitemOpen
  \bibfield  {author} {\bibinfo {author} {\bibfnamefont {T.~A.}\ \bibnamefont
  {Pasquini}}, \bibinfo {author} {\bibfnamefont {Y.}~\bibnamefont {Shin}},
  \bibinfo {author} {\bibfnamefont {C.}~\bibnamefont {Sanner}}, \bibinfo
  {author} {\bibfnamefont {M.}~\bibnamefont {Saba}}, \bibinfo {author}
  {\bibfnamefont {A.}~\bibnamefont {Schirotzek}}, \bibinfo {author}
  {\bibfnamefont {D.~E.}\ \bibnamefont {Pritchard}}, \ and\ \bibinfo {author}
  {\bibfnamefont {W.}~\bibnamefont {Ketterle}},\ }\Doi
  {\href{10.1103/PhysRevLett.93.223201}{10.1103/PhysRevLett.93.223201}}
  {\bibfield  {journal} {\bibinfo  {journal} {{Phys. Rev. Lett.}},\ }\textbf
  {\bibinfo {volume} {{93}}},\ \bibinfo {pages} {223201} (\bibinfo {year}
  {2004})}\BibitemShut {NoStop}%
\bibitem [{\citenamefont {Oberst}\ \emph {et~al.}(2005)\citenamefont {Oberst},
  \citenamefont {Kouznetsov}, \citenamefont {Shimizu}, \citenamefont {Fujita},\
  and\ \citenamefont {Shimizu}}]{Oberst2005}%
  \BibitemOpen
  \bibfield  {author} {\bibinfo {author} {\bibfnamefont {H.}~\bibnamefont
  {Oberst}}, \bibinfo {author} {\bibfnamefont {D.}~\bibnamefont {Kouznetsov}},
  \bibinfo {author} {\bibfnamefont {K.}~\bibnamefont {Shimizu}}, \bibinfo
  {author} {\bibfnamefont {J.-i.}\ \bibnamefont {Fujita}}, \ and\ \bibinfo
  {author} {\bibfnamefont {F.}~\bibnamefont {Shimizu}},\ }\Doi
  {\href{10.1103/PhysRevLett.94.013203}{10.1103/PhysRevLett.94.013203}}
  {\bibfield  {journal} {\bibinfo  {journal} {{Phys. Rev. Lett.}},\ }\textbf
  {\bibinfo {volume} {{94}}},\ \bibinfo {pages} {013203} (\bibinfo {year}
  {2005})}\BibitemShut {NoStop}%
\bibitem [{\citenamefont {Pasquini}\ \emph {et~al.}(2006)\citenamefont
  {Pasquini}, \citenamefont {Saba}, \citenamefont {Jo}, \citenamefont {Shin},
  \citenamefont {Ketterle}, \citenamefont {Pritchard}, \citenamefont {Savas},\
  and\ \citenamefont {Mulders}}]{Pasquini2006}%
  \BibitemOpen
  \bibfield  {author} {\bibinfo {author} {\bibfnamefont {T.~A.}\ \bibnamefont
  {Pasquini}}, \bibinfo {author} {\bibfnamefont {M.}~\bibnamefont {Saba}},
  \bibinfo {author} {\bibfnamefont {G.-B.}\ \bibnamefont {Jo}}, \bibinfo
  {author} {\bibfnamefont {Y.}~\bibnamefont {Shin}}, \bibinfo {author}
  {\bibfnamefont {W.}~\bibnamefont {Ketterle}}, \bibinfo {author}
  {\bibfnamefont {D.~E.}\ \bibnamefont {Pritchard}}, \bibinfo {author}
  {\bibfnamefont {T.~A.}\ \bibnamefont {Savas}}, \ and\ \bibinfo {author}
  {\bibfnamefont {N.}~\bibnamefont {Mulders}},\ }\Doi
  {\href{10.1103/PhysRevLett.97.093201}{10.1103/PhysRevLett.97.093201}}
  {\bibfield  {journal} {\bibinfo  {journal} {{Phys. Rev. Lett.}},\ }\textbf
  {\bibinfo {volume} {{97}}},\ \bibinfo {pages} {093201} (\bibinfo {year}
  {2006})}\BibitemShut {NoStop}%
\bibitem [{\citenamefont {Zhao}\ \emph {et~al.}(2010)\citenamefont {Zhao},
  \citenamefont {Schewe}, \citenamefont {Meijer},\ and\ \citenamefont
  {Schoellkopf}}]{Zhao2010}%
  \BibitemOpen
  \bibfield  {author} {\bibinfo {author} {\bibfnamefont {B.~S.}\ \bibnamefont
  {Zhao}}, \bibinfo {author} {\bibfnamefont {H.~C.}\ \bibnamefont {Schewe}},
  \bibinfo {author} {\bibfnamefont {G.}~\bibnamefont {Meijer}}, \ and\ \bibinfo
  {author} {\bibfnamefont {W.}~\bibnamefont {Schoellkopf}},\ }\Doi
  {\href{10.1103/PhysRevLett.105.133203}{10.1103/PhysRevLett.105.133203}}
  {\bibfield  {journal} {\bibinfo  {journal} {{Phys. Rev. Lett.}},\ }\textbf
  {\bibinfo {volume} {{105}}},\ \bibinfo {pages} {133203} (\bibinfo {year}
  {2010})}\BibitemShut {NoStop}%
\bibitem [{\citenamefont {Zhao}\ \emph {et~al.}(2011)\citenamefont {Zhao},
  \citenamefont {Meijer},\ and\ \citenamefont {Schoellkopf}}]{Zhao2011}%
  \BibitemOpen
  \bibfield  {author} {\bibinfo {author} {\bibfnamefont {B.~S.}\ \bibnamefont
  {Zhao}}, \bibinfo {author} {\bibfnamefont {G.}~\bibnamefont {Meijer}}, \ and\
  \bibinfo {author} {\bibfnamefont {W.}~\bibnamefont {Schoellkopf}},\ }\Doi
  {\href{10.1126/science.1200911}{10.1126/science.1200911}} {\bibfield
  {journal} {\bibinfo  {journal} {{Science}},\ }\textbf {\bibinfo {volume}
  {{331}}},\ \bibinfo {pages} {892} (\bibinfo {year} {2011})}\BibitemShut
  {NoStop}%
\bibitem [{\citenamefont {Berry}\ and\ \citenamefont
  {Mount}(1972)}]{Berry1972}%
  \BibitemOpen
  \bibfield  {author} {\bibinfo {author} {\bibfnamefont {M.~V.}\ \bibnamefont
  {Berry}}\ and\ \bibinfo {author} {\bibfnamefont {K.~E.}\ \bibnamefont
  {Mount}},\ }\Doi {\href{10.1088/0034-4885}{10.1088/0034-4885}} {\bibfield
  {journal} {\bibinfo  {journal} {{Rep. on Progr. in Phys.}},\ }\textbf
  {\bibinfo {volume} {{35}}},\ \bibinfo {pages} {315} (\bibinfo {year}
  {1972})}\BibitemShut {NoStop}%
\bibitem [{\citenamefont {B\"{o}heim}\ \emph {et~al.}(1982)\citenamefont
  {B\"{o}heim}, \citenamefont {Brenig},\ and\ \citenamefont
  {Stutzki}}]{Boheim1982}%
  \BibitemOpen
  \bibfield  {author} {\bibinfo {author} {\bibfnamefont {J.}~\bibnamefont
  {B\"{o}heim}}, \bibinfo {author} {\bibfnamefont {W.}~\bibnamefont {Brenig}},
  \ and\ \bibinfo {author} {\bibfnamefont {J.}~\bibnamefont {Stutzki}},\ }\Doi
  {\href{10.1007/BF02026427}{10.1007/BF02026427}} {\bibfield  {journal}
  {\bibinfo  {journal} {{Zeitschrift f\"{u}r Physik B Condensed Matter}},\
  }\textbf {\bibinfo {volume} {{48}}},\ \bibinfo {pages} {43} (\bibinfo {year}
  {1982})}\BibitemShut {NoStop}%
\bibitem [{\citenamefont {Clougherty}\ and\ \citenamefont
  {Kohn}(1992)}]{Clougherty1992}%
  \BibitemOpen
  \bibfield  {author} {\bibinfo {author} {\bibfnamefont {D.~P.}\ \bibnamefont
  {Clougherty}}\ and\ \bibinfo {author} {\bibfnamefont {W.}~\bibnamefont
  {Kohn}},\ }\Doi {\href{10.1103/PhysRevB.46.4921}{10.1103/PhysRevB.46.4921}}
  {\bibfield  {journal} {\bibinfo  {journal} {{Phys. Rev. B}},\ }\textbf
  {\bibinfo {volume} {{46}}},\ \bibinfo {pages} {4921} (\bibinfo {year}
  {1992})}\BibitemShut {NoStop}%
\bibitem [{\citenamefont {Carraro}\ and\ \citenamefont
  {Cole}(1992)}]{Carraro1992}%
  \BibitemOpen
  \bibfield  {author} {\bibinfo {author} {\bibfnamefont {C.}~\bibnamefont
  {Carraro}}\ and\ \bibinfo {author} {\bibfnamefont {M.~W.}\ \bibnamefont
  {Cole}},\ }\Doi {\href{10.1103/PhysRevB.45.12930}{10.1103/PhysRevB.45.12930}}
  {\bibfield  {journal} {\bibinfo  {journal} {{Phys. Rev. B}},\ }\textbf
  {\bibinfo {volume} {{45}}},\ \bibinfo {pages} {12930} (\bibinfo {year}
  {1992})}\BibitemShut {NoStop}%
\bibitem [{\citenamefont {Henkel}\ \emph {et~al.}(1996)\citenamefont {Henkel},
  \citenamefont {Westbrook},\ and\ \citenamefont {Aspect}}]{Henkel1996}%
  \BibitemOpen
  \bibfield  {author} {\bibinfo {author} {\bibfnamefont {C.}~\bibnamefont
  {Henkel}}, \bibinfo {author} {\bibfnamefont {C.~I.}\ \bibnamefont
  {Westbrook}}, \ and\ \bibinfo {author} {\bibfnamefont {A.}~\bibnamefont
  {Aspect}},\ }\Doi {\href{10.1364/JOSAB.13.000233}{10.1364/JOSAB.13.000233}}
  {\bibfield  {journal} {\bibinfo  {journal} {{J. Opt. Soc. Am.}},\ }\textbf
  {\bibinfo {volume} {{B13}}},\ \bibinfo {pages} {233} (\bibinfo {year}
  {1996})}\BibitemShut {NoStop}%
\bibitem [{\citenamefont {Friedrich}\ \emph {et~al.}(2002)\citenamefont
  {Friedrich}, \citenamefont {Jacoby},\ and\ \citenamefont
  {Meister}}]{Friedrich2002}%
  \BibitemOpen
  \bibfield  {author} {\bibinfo {author} {\bibfnamefont {H.}~\bibnamefont
  {Friedrich}}, \bibinfo {author} {\bibfnamefont {G.}~\bibnamefont {Jacoby}}, \
  and\ \bibinfo {author} {\bibfnamefont {C.~G.}\ \bibnamefont {Meister}},\
  }\Doi {\href{10.1103/PhysRevA.65.032902}{10.1103/PhysRevA.65.032902}}
  {\bibfield  {journal} {\bibinfo  {journal} {{Phys. Rev. A}},\ }\textbf
  {\bibinfo {volume} {{65}}},\ \bibinfo {pages} {032902} (\bibinfo {year}
  {2002})}\BibitemShut {NoStop}%
\bibitem [{\citenamefont {Friedrich}\ and\ \citenamefont
  {Jurisch}(2004)}]{Friedrich2004}%
  \BibitemOpen
  \bibfield  {author} {\bibinfo {author} {\bibfnamefont {H.}~\bibnamefont
  {Friedrich}}\ and\ \bibinfo {author} {\bibfnamefont {A.}~\bibnamefont
  {Jurisch}},\ }\Doi
  {\href{10.1103/PhysRevLett.92.103202}{10.1103/PhysRevLett.92.103202}}
  {\bibfield  {journal} {\bibinfo  {journal} {{Phys. Rev. Lett.}},\ }\textbf
  {\bibinfo {volume} {{92}}},\ \bibinfo {pages} {103202} (\bibinfo {year}
  {2004})}\BibitemShut {NoStop}%
\bibitem [{\citenamefont {Friedrich}\ and\ \citenamefont
  {Trost}(2004)}]{Friedrich2004a}%
  \BibitemOpen
  \bibfield  {author} {\bibinfo {author} {\bibfnamefont {H.}~\bibnamefont
  {Friedrich}}\ and\ \bibinfo {author} {\bibfnamefont {J.}~\bibnamefont
  {Trost}},\ }\Doi
  {\href{10.1016/j.physrep.2004.04.001}{10.1016/j.physrep.2004.04.001}}
  {\bibfield  {journal} {\bibinfo  {journal} {{Physics Reports}},\ }\textbf
  {\bibinfo {volume} {{397}}},\ \bibinfo {pages} {359} (\bibinfo {year}
  {2004})}\BibitemShut {NoStop}%
\bibitem [{\citenamefont {Voronin}\ and\ \citenamefont
  {Froelich}(2005)}]{Voronin2005}%
  \BibitemOpen
  \bibfield  {author} {\bibinfo {author} {\bibfnamefont {A.~Y.}\ \bibnamefont
  {Voronin}}\ and\ \bibinfo {author} {\bibfnamefont {P.}~\bibnamefont
  {Froelich}},\ }\href@noop {} {\bibfield  {journal} {\bibinfo  {journal}
  {{J. Phys. B}},\ }\textbf
  {\bibinfo {volume} {{38}}},\ \bibinfo {pages} {L301} (\bibinfo {year}
  {2005})}\BibitemShut {NoStop}%
\bibitem [{\citenamefont {Judd}\ \emph {et~al.}(2011)\citenamefont {Judd},
  \citenamefont {Scott}, \citenamefont {Martin}, \citenamefont {Kaczmarek},\
  and\ \citenamefont {Fromhold}}]{Judd2011}%
  \BibitemOpen
  \bibfield  {author} {\bibinfo {author} {\bibfnamefont {T.~E.}\ \bibnamefont
  {Judd}}, \bibinfo {author} {\bibfnamefont {R.~G.}\ \bibnamefont {Scott}},
  \bibinfo {author} {\bibfnamefont {A.~M.}\ \bibnamefont {Martin}}, \bibinfo
  {author} {\bibfnamefont {B.}~\bibnamefont {Kaczmarek}}, \ and\ \bibinfo
  {author} {\bibfnamefont {T.~M.}\ \bibnamefont {Fromhold}},\ }\Doi
  {\href{10.1088/1367-2630/13/8/083020}{10.1088/1367-2630/13/8/083020}}
  {\bibfield  {journal} {\bibinfo  {journal} {{New J. of Phys.}},\
  }\textbf {\bibinfo {volume} {{13}}},\ \bibinfo {pages} {083020} (\bibinfo
  {year} {2011})}\BibitemShut {NoStop}%
\bibitem [{\citenamefont {Voronin}\ \emph {et~al.}(2012)\citenamefont
  {Voronin}, \citenamefont {Nesvizhevsky},\ and\ \citenamefont
  {Reynaud}}]{Voronin2012}%
  \BibitemOpen
  \bibfield  {author} {\bibinfo {author} {\bibfnamefont {A.~Y.}\ \bibnamefont
  {Voronin}}, \bibinfo {author} {\bibfnamefont {V.~V.}\ \bibnamefont
  {Nesvizhevsky}}, \ and\ \bibinfo {author} {\bibfnamefont {S.}~\bibnamefont
  {Reynaud}},\ }\href@noop {} {\bibfield  {journal} {\bibinfo  {journal}
  {{J. Phys. B}},\ }\textbf
  {\bibinfo {volume} {{45}}} (\bibinfo {year} {2012})}\BibitemShut {NoStop}%
\bibitem [{\citenamefont {Dufour}\ \emph
  {et~al.}(2013){\natexlab{a}}\citenamefont {Dufour}, \citenamefont
  {G\'{e}rardin}, \citenamefont {Gu\'{e}rout}, \citenamefont {Lambrecht},
  \citenamefont {Nesvizhevsky}, \citenamefont {Reynaud},\ and\ \citenamefont
  {Voronin}}]{Dufour2013qrefl}%
  \BibitemOpen
  \bibfield  {author} {\bibinfo {author} {\bibfnamefont {G.}~\bibnamefont
  {Dufour}}, \bibinfo {author} {\bibfnamefont {A.}~\bibnamefont
  {G\'{e}rardin}}, \bibinfo {author} {\bibfnamefont {R.}~\bibnamefont
  {Gu\'{e}rout}}, \bibinfo {author} {\bibfnamefont {A.}~\bibnamefont
  {Lambrecht}}, \bibinfo {author} {\bibfnamefont {V.~V.}\ \bibnamefont
  {Nesvizhevsky}}, \bibinfo {author} {\bibfnamefont {S.}~\bibnamefont
  {Reynaud}}, \ and\ \bibinfo {author} {\bibfnamefont {A.~Y.}\ \bibnamefont
  {Voronin}},\ }\Doi
  {\href{10.1103/PhysRevA.87.012901}{10.1103/PhysRevA.87.012901}} {\bibfield
  {journal} {\bibinfo  {journal} {{Phys. Rev. A}},\ }\textbf {\bibinfo {volume}
  {{87}}},\ \bibinfo {pages} {012901} (\bibinfo {year}
  {2013}{\natexlab{a}})}\BibitemShut {NoStop}%
\bibitem [{\citenamefont {Dufour}\ \emph
  {et~al.}(2013){\natexlab{b}}\citenamefont {Dufour}, \citenamefont
  {Gu\'{e}rout}, \citenamefont {Lambrecht}, \citenamefont {Nesvizhevsky},
  \citenamefont {Reynaud},\ and\ \citenamefont {Voronin}}]{Dufour2013porous}%
  \BibitemOpen
  \bibfield  {author} {\bibinfo {author} {\bibfnamefont {G.}~\bibnamefont
  {Dufour}}, \bibinfo {author} {\bibfnamefont {R.}~\bibnamefont {Gu\'{e}rout}},
  \bibinfo {author} {\bibfnamefont {A.}~\bibnamefont {Lambrecht}}, \bibinfo
  {author} {\bibfnamefont {V.~V.}\ \bibnamefont {Nesvizhevsky}}, \bibinfo
  {author} {\bibfnamefont {S.}~\bibnamefont {Reynaud}}, \ and\ \bibinfo
  {author} {\bibfnamefont {A.~Y.}\ \bibnamefont {Voronin}},\ }\Doi
  {\href{10.1103/PhysRevA.87.022506}{10.1103/PhysRevA.87.022506}} {\bibfield
  {journal} {\bibinfo  {journal} {{Phys. Rev. A}},\ }\textbf {\bibinfo {volume}
  {{87}}},\ \bibinfo {pages} {022506} (\bibinfo {year}
  {2013}{\natexlab{b}})}\BibitemShut {NoStop}%
\bibitem [{\citenamefont {Liouville}(1836)}]{Liouville1836}%
  \BibitemOpen
  \bibfield  {author} {\bibinfo {author} {\bibfnamefont {J.}~\bibnamefont
  {Liouville}},\ }\href
  {\href{http://gallica.bnf.fr/ark:/12148/bpt6k16380x/f259n13.capture}{http://%
gallica.bnf.fr/ark:/12148/bpt6k16380x/f259n13.capture}} {\bibfield  {journal}
  {\bibinfo  {journal} {{J. de math. pures et
  appl.}},\ }\textbf {\bibinfo {volume} {{1}}},\ \bibinfo {pages}
  {253} (\bibinfo {year} {1836})}\BibitemShut {NoStop}%
\bibitem [{\citenamefont {Liouville}(1837)}]{Liouville1837}%
  \BibitemOpen
  \bibfield  {author} {\bibinfo {author} {\bibfnamefont {J.}~\bibnamefont
  {Liouville}},\ }\href
  {\href{http://gallica.bnf.fr/ark:/12148/cb343487840/date}{http://gallica.bnf%
.fr/ark:/12148/cb343487840/date}} {\bibfield  {journal} {\bibinfo  {journal}
  {{J. de math. pures et appl.}},\ }\textbf {\bibinfo
  {volume} {{2}}},\ \bibinfo {pages} {16} (\bibinfo {year} {1837})}\BibitemShut
  {NoStop}%
\bibitem [{\citenamefont {Olver}(1997)}]{Olver1997}%
  \BibitemOpen
  \bibfield  {author} {\bibinfo {author} {\bibfnamefont {F.}~\bibnamefont
  {Olver}},\ }\href@noop {} {{\emph {\bibinfo {title}
  {{Asymptotics and Special Functions}}}}}\ (\bibinfo  {publisher} {{Taylor \&
  Francis}},\ \bibinfo {year} {1997})\ ISBN \bibinfo {isbn}
  {9781568810690}\BibitemShut {NoStop}%
\bibitem [{\citenamefont {Milson}(1998)}]{Milson1998}%
  \BibitemOpen
  \bibfield  {author} {\bibinfo {author} {\bibfnamefont {R.}~\bibnamefont
  {Milson}},\ }\Doi {\href{10.1023/A:1026696709617}{10.1023/A:1026696709617}}
  {\bibfield  {journal} {\bibinfo  {journal} {{Int. J. of
  Theor. Phys.}},\ }\textbf {\bibinfo {volume} {{37}}},\ \bibinfo
  {pages} {1735} (\bibinfo {year} {1998})}\BibitemShut {NoStop}%
\bibitem [{\citenamefont {Whitton}\ and\ \citenamefont
  {Connor}(1973)}]{Whitton1973}%
  \BibitemOpen
  \bibfield  {author} {\bibinfo {author} {\bibfnamefont {W.}~\bibnamefont
  {Whitton}}\ and\ \bibinfo {author} {\bibfnamefont {J.}~\bibnamefont
  {Connor}},\ }\Doi
  {\href{10.1080/00268977300102661}{10.1080/00268977300102661}} {\bibfield
  {journal} {\bibinfo  {journal} {{Molec. Phys.}},\ }\textbf {\bibinfo
  {volume} {{26}}},\ \bibinfo {pages} {1511} (\bibinfo {year}
  {1973})}\BibitemShut {NoStop}%
\end{thebibliography}%

\end{document}